\documentstyle[preprint,aps,epsf,floats]{revtex}
\begin{document}

\draft

{\tighten

\preprint{\vbox{\hbox{CALT-68-2110}\hbox{JHU-TIPAC-97009}
		\hbox{hep-ph/9705235} \hbox{} }}

\title{$V_{ub}$ from the Hadronic Invariant Mass Spectrum \\
  in Semileptonic $B$ Decay}

\author{Adam F.\ Falk$\,^a$, Zoltan Ligeti$\,^b$ and Mark B.\ Wise$\,^b$ }

\address{ \vbox{\vskip 0.truecm}
  $^a$Department of Physics and Astronomy, The Johns Hopkins University \\
    3400 North Charles Street, Baltimore, MD 21218 \\[6pt]
  $^b$California Institute of Technology, Pasadena, CA 91125 }

\maketitle

\begin{abstract}%
The hadronic invariant mass spectrum for the inclusive charmless semileptonic decay $B\to
X_u\,e\,\bar\nu_e$ is studied.  Particular attention is paid to the region
$s_H<m_D^2$, which may be useful for extracting the value of $|V_{ub}|$.
The sensitivity of the spectrum to the parameter $\bar\Lambda \equiv m_B - m_b$ is explored. 
Perturbative QCD corrections to ${\rm d}\Gamma/{\rm d}s_H$ of order
$\alpha_s^2\beta_0$ are calculated.  For $s_H\sim\bar\Lambda\,m_b$
nonperturbative QCD effects are important and the shape of the invariant mass
spectrum is controlled by the $B$ meson matrix element of an infinite sum of
local operators.  The utility of the hadronic mass spectrum for extracting
$|V_{ub}|$ is explored.

\end{abstract}

}
\newpage

The traditional method for extracting $|V_{ub}|$ from experimental data
involves a study of the electron energy spectrum in inclusive charmless semileptonic $B$
decay~\cite{CLEO1}.  For a particular hadronic final state $X$ the maximum
electron energy is $E_e^{\rm (max)}=(m_B^2-m_X^2)/2m_B$ (in the $B$ rest frame),
and consequently electrons with energies in the endpoint region
$m_B/2>E_e>(m_B^2-m_D^2)/2m_B$ (neglecting the pion mass) must arise from the
$b \to u$ transition.  A determination of $|V_{ub}|$ from experimental data on
the electron spectrum in the endpoint region is possible, provided a
theoretical prediction for the electron spectrum can be made.

Recently there has been considerable theoretical progress in our understanding
of inclusive semileptonic $B$ decay~\cite{CGG,incl,MaWi}.  It is based on the
use of the operator product expansion (OPE) and heavy quark effective theory
(HQET) to include in the differential decay rate nonperturbative effects
suppressed by powers of $\Lambda_{\rm QCD}/m_b$.  At leading order in the
$\Lambda_{\rm QCD}/m_b$ expansion the $B$ meson decay rate is equal to the
$b$ quark decay rate.  There are no nonperturbative corrections of order
$\Lambda_{\rm QCD}/m_b$.  At order $\Lambda_{\rm QCD}^2/m_b^2$~\cite{incl,MaWi}
the nonperturbative corrections are characterized by two HQET matrix elements
$\lambda_{1,2}$, which are defined by
\begin{eqnarray}\label{1}
\lambda_1 &=& \langle B(v)|\, \bar h_v^{(b)}\, (iD)^2\, h_v^{(b)}\,
  |B(v)\rangle/2m_B \,, \nonumber\\*
\lambda_2 &=& \langle B(v)|\, {g_s\over 2}\, \bar h_v^{(b)}\,
  \sigma_{\mu\nu} G^{\mu\nu}\, h_v^{(b)}\, | B(v) \rangle/6m_B \,.
\end{eqnarray}
These matrix elements also occur in the expansion of the $B$ and $B^*$
masses in powers of $\Lambda_{\rm QCD}/m_b$,
\begin{eqnarray}\label{3}
m_B &=& m_b + \bar\Lambda - (\lambda_1 + 3 \lambda_2)/2m_b + \ldots \,,
  \nonumber\\*
m_{B^*} &=& m_b + \bar\Lambda - (\lambda_1 - \lambda_2)/2m_b + \ldots \,.
\end{eqnarray}
Similar formulae hold for the $D$ and $D^*$ masses.  The parameters $\lambda_1$
and $\lambda_2$ are independent of the heavy $b$ quark mass (there is a weak
logarithmic dependence in $\lambda_2$) and are of order $\Lambda_{\rm QCD}^2$. 
The measured $B^*-B$ mass splitting fixes $\lambda_2=0.12\,{\rm GeV}^2$.  The
mass formulae for the $B$ and $B^*$ mesons involve not only $\lambda_{1,2}$ but
also a parameter $\bar\Lambda$, which is the difference between the $B$ meson
mass and the $b$ quark mass in the $m_b\to\infty$ limit.  The measured $B$
semileptonic decay spectrum in the region $E_e \geq 1.5\,$GeV has been used to
determine $\bar\Lambda \simeq 0.4\,$GeV and $\lambda_1\simeq-0.2\,{\rm GeV}^2$
\cite{gremm}.  Unfortunately the uncertainties from terms of order
$(\Lambda_{\rm QCD}/m_b)^3$ are quite large~\cite{AM}.  (A linear
combination of $\bar\Lambda$ and $\lambda_1$ is rather well constrained, but the
individual values are more uncertain.)

The maximum electron energy in semileptonic $b$ quark decay is $m_b/2$.  This
is less than the physical endpoint by $\bar\Lambda/2$, which is comparable in
size to the endpoint region $\Delta
E_e^{\rm(endpoint)}=m_D^2/2m_B\simeq0.33\,$GeV.  Using the operator product
expansion and HQET, the effects which extend the electron spectrum beyond its
partonic value appear as singular terms in the prediction for ${\rm
d}\Gamma/{\rm d}E_e$ involving derivatives of delta functions, $\delta^{(n)}
(E_e-m_b/2)$.  Near the endpoint the electron spectrum must be smeared over a
region of energies $\Delta E_e$ before theory can be compared with
experiment.  If the smearing region $\Delta E_e$ is much smaller than
$\Lambda_{\rm QCD}$, then higher dimension operators in the OPE become
successively more important and the OPE is not useful for describing the
electron energy spectrum.  For $\Delta E_e$ much greater than $\Lambda_{\rm
QCD}$, higher dimension operators become successively less important and a
useful prediction for the electron spectrum can be made using the first few
terms in the OPE.  When $\Delta E_e\sim\Lambda_{\rm QCD}$ there is an infinite
series of terms in the OPE which are all equally important.  Since $\Delta
E_e^{\rm(endpoint)}$ is about $\Lambda_{\rm QCD}$, it seems unlikely that
predictions based on a few low dimension operators in the OPE can successfully
determine the electron spectrum in this region.

In the future, another possibility for determining $|V_{ub}|$ may come from a
comparison of the measured hadronic invariant mass spectrum in the region
$s_H<m_D^2$ with theoretical predictions.  Here $s_H=(p_B-q)^2$, where $p_B$ is
the $B$ meson four-momentum, and $q=p_e+p_{\bar\nu_e}$ is the sum of the lepton
four-momenta.  An obvious advantage to studying this quantity rather than the
lepton energy spectrum is that most of the $B\to X_u\,e\,\bar\nu_e$ decays are
expected to lie in the region $s_H<m_D^2$, while only a small fraction of the
$B\to X_u\,e\,\bar\nu_e$ decays have electron energies in the endpoint region. 
Both the invariant mass region, $s_H<m_D^2$, and the electron endpoint region,
$m_B/2>E_e>(m_B^2-m_D^2)/2m_B$, receive contributions from hadronic
final states with invariant masses that range up to $m_D$.  However, for the
electron endpoint region the contribution of the states with masses nearer to
$m_D$ is kinematically suppressed since they typically decay to lower energy
electrons.  In fact, in the ISGW model~\cite{ISGW} the electron endpoint region
is dominated by the $\pi$ and the $\rho$, with higher mass states making only a
small contribution.  The situation is very different for the low invariant mass
region, $s_H<m_D^2$, with no cut on the electron energy.  Now all states with
invariant masses up to $m_D$ contribute without any preferential weighting
towards the lowest mass ones.  In the ISGW model the $\pi$ and the $\rho$
mesons comprise only about a quarter of the $B$ semileptonic decays to states
with masses less than $m_D$.  Consequently, it is much more likely that the
first few terms in the OPE will provide an accurate description $B$
semileptonic decay in the region $s_H<m_D^2$ than in the endpoint region of the
electron energy spectrum.  Combining the invariant mass constraint,
$s_H<m_D^2$, with a modest cut on the electron energy will not destroy this
conclusion.  (Such a cut will probably be required experimentally for the
direct measurement of $s_H$ via the neutrino reconstruction technique.)  We
also expect that the $B\to X_u\,e\,\bar\nu_e$ rate in the invariant mass region
$s_H<m_D^2$ is less sensitive to nonperturbative effects than is the rate in the
hadron energy region $E_H<m_D$ (in the $B$ rest frame)~\cite{energy}, since the
hadron energy constraint cuts out more of the phase space for states with mass
near $m_D$ than for the lower mass states.  In this letter we explore the
utility of the hadronic invariant mass spectrum \cite{mass} for determining the
magnitude of $V_{ub}$.  The possibility of using the hadronic invariant mass
spectrum in $B\to X_c\,e\,\bar\nu_e$ to determine $\bar\Lambda$ and $\lambda_1$
was discussed in Ref.~\cite{FLS}.  The technique is promising but awaits better
data on ${\rm d}\Gamma/{\rm d}s_H$.

To begin with, consider the contribution of dimension three operators in the
OPE to the hadronic mass squared spectrum in $B\to X_u\,e\,\bar\nu_e$ decay. 
This is equivalent to $b$ quark decay and implies a result for ${\rm
d}\Gamma/{\rm d}E_0\,{\rm d}s_0$ (where $E_0=p_b\cdot(p_b-q)/m_b$ and
$s_0=(p_b-q)^2$ are the energy and invariant mass of the strongly interacting
partons arising from the $b$ quark decay) that can easily be calculated using
perturbative QCD up to order $\alpha_s^2\beta_0$.  Even at this leading order
in the OPE there are important nonperturbative effects that come from the
relation between the $b$ quark mass and the $B$ meson mass in Eqs.~(\ref{3}). 
The most significant effect comes from $\bar\Lambda$, and including only it
({\it i.e.}, neglecting the effect of $\lambda_{1,2}$), the hadronic invariant
mass $s_H$ is related to $s_0$ and $E_0$ by~\cite{FLS}
\begin{equation}\label{5}
s_H = s_0 + 2 \bar\Lambda E_0 + \bar\Lambda^2 \,.
\end{equation}
Changing variables from $(s_0, E_0)$ to $(s_H, E_0)$ and integrating $E_0$
over the range
\begin{equation}\label{6}
\sqrt{s_H} - \bar\Lambda < E_0 <
  {1\over 2m_B}\, (s_H - 2\bar\Lambda m_B + m_B^2),
\end{equation}
gives ${\rm d}\Gamma/{\rm d}s_H$, where $\bar\Lambda^2<s_H<m_B^2$.  Feynman
diagrams with only a $u$-quark in the final state contribute at $s_0=0$, which
corresponds to the region $\bar\Lambda^2<s_H<\bar\Lambda m_B$.

Although ${\rm d}\Gamma/{\rm d}s_H$ is integrable in perturbation theory, it
has a double logarithmic singularity at $s_H=\bar\Lambda m_B$.  At higher
orders in perturbation theory, increasing powers of
$\alpha_s\ln^2[(s_H-\bar\Lambda m_B)/m_B^2]$ appear in the invariant mass
spectrum.\footnote{For recent discussions of a similar phenomenon in the
electron energy spectrum, see Ref.~\cite{Ira}.} Therefore, ${\rm d}\Gamma/{\rm
d}s_H$ in the vicinity of $s_H=\bar\Lambda m_B$ is hard to predict reliably
even in perturbation theory.  (In the region $s_H\lesssim\bar\Lambda m_B$
nonperturbative effects, which we discuss later, are also important.)  The
behavior of the spectrum near $s_H=\bar\Lambda m_B$ becomes less important for
observables that average over larger regions of the spectrum, such as ${\rm
d}\Gamma/{\rm d}s_H$ integrated over $s_H<\Delta^2$, with $\Delta^2$
significantly greater than $\bar\Lambda m_B$.  Therefore, we present results
for ${\rm d}\Gamma/{\rm d}s_H$ in the region $s_H>\bar\Lambda m_B$, where only
the bremsstrahlung Feynman diagrams contribute.  Calculating these Feynman
diagrams gives the differential decay rate
\begin{eqnarray}\label{8}
{{\rm d}\Gamma(B\to X_u\,e\,\bar\nu_e)\over {\rm d}s_H} &=&
  {G_F^2\, m_B^3\over 192\pi^3}\, |V_{ub}|^2
  \bigg(1-{\bar\Lambda\over m_B}\bigg)^3 \nonumber\\*
&\times& \bigg[ {\alpha_s(\sqrt{s_H})\over\pi}\, X(s_H,\bar\Lambda)
  + \bigg({\alpha_s(\sqrt{s_H})\over\pi}\bigg)^2\, \beta_0\, Y(s_H,\bar\Lambda)
  + \ldots \bigg] \,,
\end{eqnarray}
where $\beta_0=11-2n_f/3$ is the one-loop beta function of QCD.

\begin{figure}[t]
\centerline{\epsfysize=8truecm \epsfbox{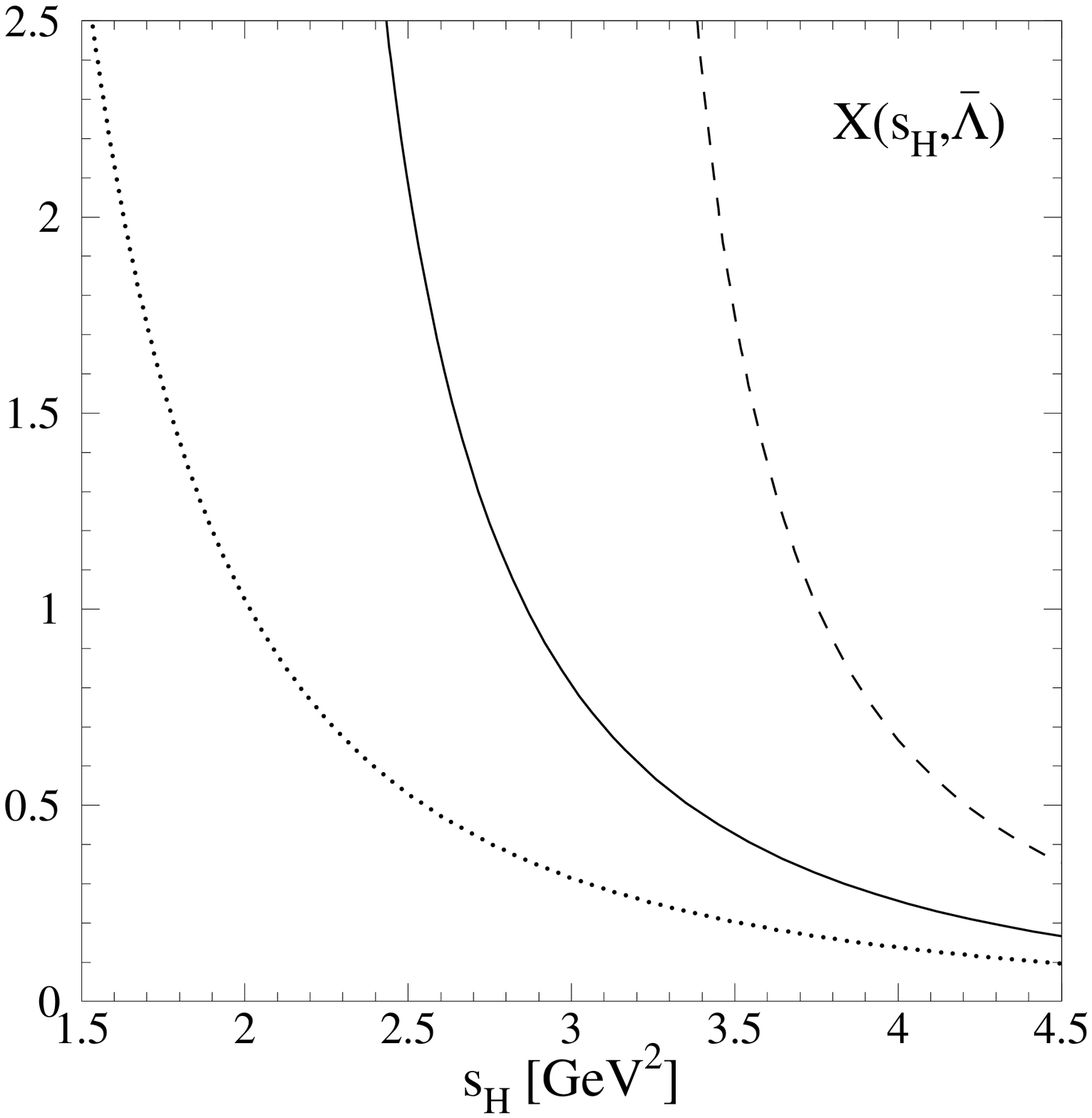}
  \epsfysize=8truecm \epsfbox{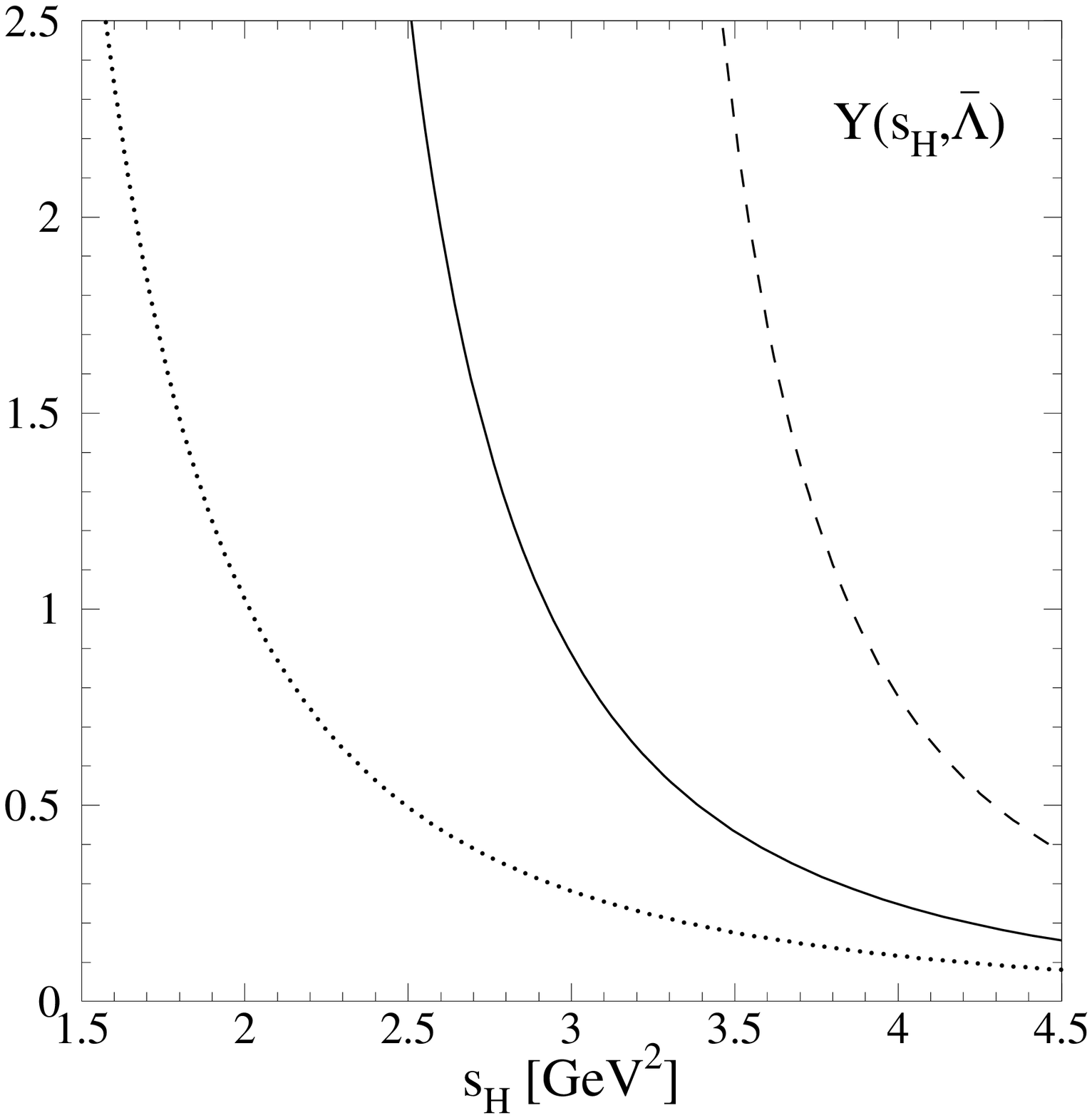}}
\caption[1]{The functions $X(s_H,\bar\Lambda)$ and $Y(s_H,\bar\Lambda)$
defined in Eq.~(\ref{8}) for $\bar\Lambda=0.2\,$GeV (dotted curve),
$0.4\,$GeV (solid curve), and $0.6\,$GeV (dashed curve).}
\end{figure}

In Figs.~1 we plot $X(s_H,\bar\Lambda)$ and $Y(s_H,\bar\Lambda)$ as
functions of $s_H$ for $\bar\Lambda=0.2$, $0.4$ and $0.6\,$GeV.  The
$\overline{\rm MS}$ scheme is used for the strong coupling, and we choose to
evaluate $\alpha_s$ at the scale $\sqrt{s_H}$.  While $Y(s_H,\bar\Lambda)$ is
sensitive to this choice, the sum of the two terms in the square brackets in
Eq.~(\ref{8}) has only a weak scale-dependence.  Even though the
$\alpha_s^2\beta_0$ correction is as large as the $\alpha_s$ term, this does
not necessarily imply a problem with the perturbative corrections, since there
is a renormalon ambiguity of order $\Lambda_{\rm QCD}$ in $\bar\Lambda$ which
cancels a renormalon ambiguity in the perturbative QCD corrections.

\begin{figure}[t]
\centerline{\epsfysize=9truecm \epsfbox{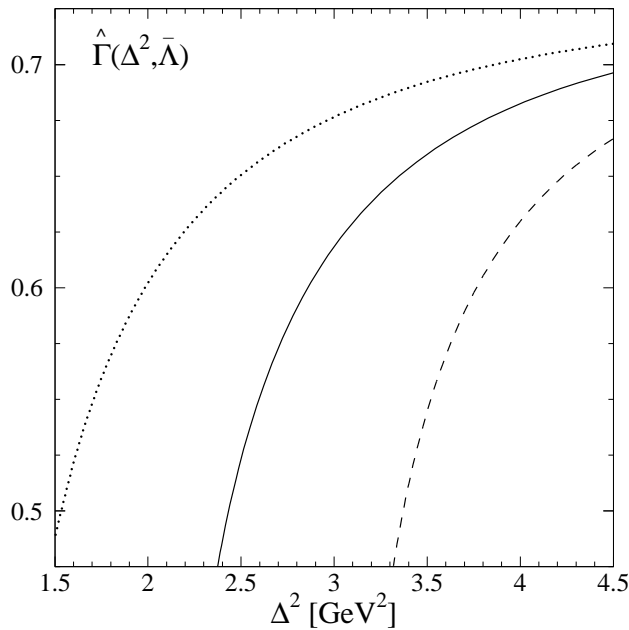}}
\caption[2]{The function $\hat\Gamma(\Delta^2,\bar\Lambda)$ defined in
Eq.~(\ref{9}) as a function of $\Delta^2$ for $\bar\Lambda=0.2\,$GeV
(dotted curve), $0.4\,$GeV (solid curve), and $0.6\,$GeV (dashed curve).}
\end{figure}

To examine the sensitivity to $\bar\Lambda$ of an extracted value of
$|V_{ub}|$ from the number
of events in a region $s_H<\Delta^2$, we define the dimensionless quantity
$\hat\Gamma(\Delta^2,\bar\Lambda)$ by
\begin{equation}\label{9}
\int_0^{\Delta^2} {\rm d}s_H\,
  {{\rm d}\Gamma(B\to X_u\,e\,\bar\nu_e)\over {\rm d}s_H} =
  {G_F^2\,m_B^5\over 192\pi^3}\, |V_{ub}|^2\,
  \bigg(1-{\bar\Lambda\over m_B}\bigg)^5\, \hat\Gamma(\Delta^2,\bar\Lambda) \,.
\end{equation}
In Fig.~2 we plot $\hat\Gamma(\Delta^2,\bar\Lambda)$ as a function of
$\Delta^2$ for $\bar\Lambda=0.2$, $0.4$ and $0.6\,$GeV in the region
$\bar\Lambda m_B<\Delta^2<4.5\,{\rm GeV}^2$, using $\alpha_s(m_b)=0.2$.  These
curves approach $\hat\Gamma(m_B^2,\bar\Lambda)\simeq0.73$ as $\Delta^2\to
m_B^2$~\cite{LSV}.  The spread of the curves in Fig.~2 together with the
$(1-\bar\Lambda/m_B)^5$ dependence factored out in Eq.~(\ref{9}) suggest that
an accurate value of $|V_{ub}|$ can be obtained from the number of events in a
region $s_H<\Delta^2$ if $\Delta^2$ is not much below $m_D^2=3.5\,{\rm GeV}^2$,
and if a reasonably precise determination of $\bar\Lambda$ is available.  For
example, with $\Delta^2=3.5\,{\rm GeV}^2$ and $\bar\Lambda=0.4\pm0.1\,$GeV, the
uncertainty arising from the error in $\bar\Lambda$ in the extracted value of
$|V_{ub}|$ is only 8\%.  So far nonperturbative corrections from higher
dimension operators in the OPE have been neglected.  We discuss their influence
on the extraction of $|V_{ub}|$ later.

Experimental uncertainties will cause some of the $B\to X_c\,e\,\bar\nu_e$
events to appear to have $s_H<m_D^2$.  If experimental $s_H$ resolution forces
$\Delta^2$ to be much below $m_D^2$, the uncertainties increase significantly. 
It is not clear at the present time how, for example, $\Delta^2=(1.5\,{\rm
GeV})^2$ compares to $\bar\Lambda m_B$.  For such a small value of $\Delta^2$,
our results for $\hat\Gamma(\Delta^2,\bar\Lambda)$ in Fig.~2 are only reliable
if $\bar\Lambda$ has a small value, below $0.4\,$GeV.  With
$\bar\Lambda=0.4\,$GeV, one should worry about the reliability of an extraction
of $|V_{ub}|$ based on $\Delta^2=(1.5\,{\rm GeV})^2$, since higher order
perturbative corrections and nonperturbative effects (which we discuss next) are
likely to be important.\footnote{For example, the order $\alpha_s^2\beta_0$
result predicts for $\bar\Lambda=0.4\,$GeV that a large fraction (about 40\%)
of the $B\to X_u\,e\,\bar\nu_e$ events have $s_H>(1.5\,{\rm GeV})^2$.  Taking
Fig.~2 literally and assuming $\bar\Lambda=0.3\pm0.1\,$GeV, the uncertainty in
$|V_{ub}|$ would be 17\%, but the sensitivity to uncalculated higher order
perturbative and nonperturbative effects could be significant.}

In the low mass region $s_H\lesssim\bar\Lambda m_B$, nonperturbative corrections
from higher dimension operators in the OPE are very important.  Just as in the
case of the electron spectrum in the endpoint region~\cite{shape}, the most
singular terms can be identified and summed into a shape function, $S(s_H)$. 
Neglecting perturbative QCD corrections, we write
\begin{equation}\label{10}
{{\rm d}\Gamma\over {\rm d}s_H} = {G_F^2\,m_b^5\over192\pi^3}\,
  |V_{ub}|^2\, S(s_H)\,.
\end{equation}
It is convenient to introduce the scaled variable $y=s_H/\bar\Lambda m_b$ and
define a dimensionless shape function $\hat S(y)=\bar\Lambda m_b\,S(s_H)$.  
Then
\begin{equation}\label{11}
\hat S(y) = \sum_{n=0}^\infty (-1)^n {2 A_n\over n!\,\bar\Lambda^n}\,
  {{\rm d}^n\over {\rm d}y^n} \left[ y^{n+2} \left(3 - 2y\right)
  \theta(1-y) \right] .
\end{equation}
The matrix elements $A_n$ are the same ones that determine the shape functions 
for the semileptonic $B$ decay electron energy spectrum in the endpoint region 
and the endpoint photon energy region in weak radiative $B$ decay.
Explicitly,
\begin{eqnarray}\label{12}
\langle B(v)|\, \bar h_v^{(b)}\, iD_{\mu_1} \ldots iD_{\mu_n}\, h_v^{(b)}\,
  |B(v)\rangle/2m_B &=& A_n\, v_{\mu_1} \ldots v_{\mu_n} \nonumber\\*
&+& {\rm terms~involving~the~metric~tensor} \,.
\end{eqnarray}
The $A_n$'s have dimension of $[mass]^n$, and hence the coefficients
$A_n/\bar\Lambda^n$ are dimensionless numbers of order one.  The first few
$A_n$'s are $A_0=1$, $A_1=0$, $A_2=-\lambda_1/3$, and $A_3=-\rho_1/3$.  Using
the equations of motion, $\rho_1$ can be related to the matrix element of a
four-quark operator.  In the vacuum saturation approximation,
$\rho_1=(2\pi\alpha_s/9)m_Bf_B^2$ \cite{gremm,rho1}.  Unfortunately, the
scale-dependence of this result leaves the value of $\rho_1$ highly
uncertain~\cite{AM}.

The shape function $\hat S(y)$ is an infinite sum of singular terms which gives
an invariant mass spectrum that leaks out beyond $y=1$ ({\it i.e.}, $s_H=\bar\Lambda
m_b$).  For $y\sim1$ ({\it i.e.}, $s_H\sim\bar\Lambda m_b$) all terms in
Eq.~(\ref{11}) are formally of equal importance.  Since $\bar\Lambda
m_b\approx2\,{\rm GeV}^2$ is not too far from $m_D^2$, it is necessary to
estimate the influence of the nonperturbative effects on the fraction of $B$
decays with invariant hadronic mass squared less than $m_D^2$.  It is difficult
to obtain a model-independent estimate of the leakage of events above an
experimental cutoff $s_H=\Delta^2$, given that we can estimate only the first
few moments, $A_n$.  Upper bounds on this leakage can be obtained if $\hat
S(y)$ is assumed to be positive definite.  This is consistent with the naive
interpretation of the leading singularities as constituting a nonperturbative
smearing of the rate beyond the $b$ quark decay endpoint~\cite{shape}; however,
there is no proof that this property actually holds.  The differential decay
rate is positive, but for $s_H$ comparable with $m_b^2$ there are other
nonperturbative terms which are equally important.  Furthermore, perturbative
QCD corrections have been neglected.  Models, such as the ACCMM
model~\cite{ACCMM}, do give a positive shape function.  

The fraction of events with $s_H<\Delta^2$ is given by
\begin{equation}
  F(\Delta) = \int_0^{\epsilon(\Delta)} {\rm d}y\, \hat S(y)\,,
\end{equation}
where $\epsilon(\Delta)=\Delta^2/\bar\Lambda m_b$.  Recall that the kinematic
point $s_H=\Delta^2$ corresponds to $y=\epsilon$.  Assuming a positive shape
function, $F(\Delta)$ is greater than 
\begin{equation}
F_P(\Delta) = \int_0^{m_b/\bar\Lambda} {\rm d}y\, 
  P(y,\epsilon(\Delta))\, \hat S(y)\,,
\end{equation}
provided $P(y,\epsilon)$ satisfies the following properties: ({\it i})
$P(y,\epsilon)<1$ for $y<\epsilon$; ({\it ii}) $P(y,\epsilon)<0$ for
$y>\epsilon$.  The lower bound $F(\Delta)>F_P(\Delta)$ holds for any such
$P(y,\epsilon)$.  Furthermore, if $P(y,\epsilon)=P_k(y,\epsilon)$ is a
polynomial of degree $k$ in $y$, then only the first $k$ moments, $A_k$, appear
in the bound.  Setting $P_k(y,\epsilon)=\sum_0^k a_\ell(\epsilon)\,y^\ell$, and
integrating by parts $n$ times yields
\begin{equation}
  F_{P_k}(\Delta) = \sum_{\ell=0}^k \sum_{n=0}^\ell 
  a_\ell(\epsilon)\, {A_n\over\bar\Lambda^n}\, {\ell\choose n}\,
  {2(\ell+6)\over(\ell+3)(\ell+4)}\,.
\end{equation}

As an illustration of the utility of this bound, consider first the simple 
quadratic polynomial $P_2(y)=1-y^2/\epsilon^2$.  This leads to the lower bound
\begin{equation}
  F(\Delta) > 1 - {8\over15\epsilon^2} + 
  {8\lambda_1\over45\epsilon^2\bar\Lambda^2}\,, \qquad 
  \epsilon=\Delta^2/\bar\Lambda m_b \,.
\end{equation}
For $\bar\Lambda=0.4\,$GeV and $\lambda_1=-0.2\,{\rm GeV}^2$, the bound is  
$F(m_D)>76\%$.  With larger $\bar\Lambda$, the bound weakens  
dramatically.  For $\bar\Lambda=0.6\,$GeV and $\lambda_1=-0.2\,{\rm GeV}^2$, 
it is only $F(m_D)>59\%$.  Once again, an independent determination of  
$\bar\Lambda$ and $\lambda_1$ is necessary for these bounds to become useful.
For a cubic polynomial, we also need to know $\rho_1$.  For example, 
consider $P_3(y)=1-y^3/\epsilon^3$.  Then the lower bound is
\begin{equation}
F(\Delta) > 1 - {3\over7\epsilon^3} + {3\lambda_1\over7\epsilon^3\bar\Lambda^2}
  + {\rho_1\over7\epsilon^3\bar\Lambda^3}\,, \qquad 
  \epsilon=\Delta^2/\bar\Lambda m_b \,.
\end{equation}
For $\bar\Lambda=0.4\,$GeV and $\lambda_1=-0.2\,{\rm GeV}^2$, $F(m_D)>83\%$ if
$\rho_1=0$ and $F(m_D)>87\%$ if $\rho_1=0.1\,{\rm GeV}^3$.  The dependence on
$\bar\Lambda$ is still the most important, as the latter bound falls to
$F(m_D)>69\%$ for $\bar\Lambda=0.6\,$GeV.  We could improve these bounds by
optimizing the coefficients in the polynomial $P_k$.  Since the optimization
itself will depend on $\bar\Lambda$, $\lambda_1$ and $\rho_1$, it does not seem
worth while to proceed along this line at the present time.  

If due to experimental resolution one can only use $s_H<(1.5\,{\rm GeV})^2$, then
the bounds become much weaker.  For example, using the cubic polynomial above
with $\bar\Lambda=0.4\,$GeV and $\lambda_1=-0.2\,{\rm GeV}^2$, the bound is
$F(1.5\,{\rm GeV})>36\%$ for $\rho_1=0$ and $F(1.5\,{\rm GeV})>51\%$ for
$\rho_1=0.1\,{\rm GeV}^3$.

An alternative is to resort to models for an estimate of the effect of high
order terms in the sum in Eq.~(\ref{11}).  As an example, consider the ACCMM
model~\cite{ACCMM,ACM}, where the $B$ meson is modeled by a spectator quark
with mass $m_{sp}$ and momentum $\vec p$, and a $b$ quark with momentum $-\vec
p$ and effective mass $m_b^{\rm(eff)}=m_B-\sqrt{m_{sp}^2+\vec p\,^2}$.  The
probability that the spectator quark momentum takes the value $\vec p$ is $\Phi(\vec
p\,)$.  In this model $\bar\Lambda=\int{\rm d}^3p\,\Phi(\vec
p\,)\,\sqrt{m_{sp}^2+\vec p\,^2}$, and $\lambda_1=-\int{\rm d}^3p\,\Phi(\vec
p\,)\,|\vec p\,|^2$.  

\begin{figure}[t]
\centerline{\epsfysize=9truecm \epsfbox{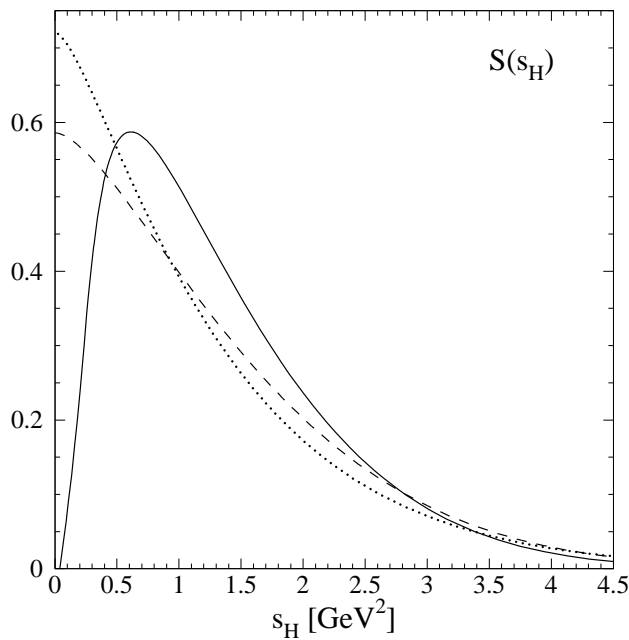}}
\caption[3]{The shape function $S(s_H)$ in the ACCMM model.  $\Phi(\vec
p\,)\propto e^{-|\vec p\,|/p_F}$ with $p_F=0.13\,$GeV and $m_{sp}=0$ (dotted
curve); $\Phi(\vec p\,)\propto e^{-|\vec p\,|^2/p_F^2}$ with $p_F=0.35\,$GeV 
and $m_{sp}=0$ (dashed curve); and $p_F=0.3\,$GeV and $m_{sp}=0.2\,$GeV (solid
curve). }
\end{figure}

To plot the shape function $S(s_H)$ in the ACCMM model, we neglect the boost
from the $b$ quark$b$ quark rest-frame into the $B$ meson rest-frame (such affects are
subleading in the $m_b\to\infty$ limit for all values of $s_H$).  In Fig.~3 we
plot the shape function for three different cases that give
$\bar\Lambda=0.4\,$GeV.  They are $\Phi(\vec p\,)\propto e^{-|\vec p\,|/p_F}$
with $p_F=0.13\,$GeV and $m_{sp}=0$ (dotted curve); $\Phi(\vec p\,)\propto
e^{-|\vec p\,|^2/p_F^2}$ with $p_F=0.35\,$GeV and $m_{sp}=0$ (dashed curve);
and $p_F=0.3\,$GeV and $m_{sp}=0.2\,$GeV (solid curve).  In these cases only
4.4\%, 4.1\% and 2.8\%, respectively, of the $B\to X_u\,e\,\bar\nu_e$ decays
have $s_H\geq m_D^2$.  Even if the leakage into the region $s_H>m_D^2$ were a
factor of two or three greater than this (a possibility which is not at all
unlikely), unknown nonperturbative effects characterized by the $A_n$'s only
give rise to about a 10\% uncertainty in the fraction of $B$ semileptonic
decays with $s_H<m_D^2$.  On the other hand, if due to charm contamination one can only use events
with $s_H<(1.5\,{\rm GeV})^2$, then the sensitivity to the shape
function is much greater.  For the three models in Fig.~3, the fraction of
$B\to X_u\,e\,\bar\nu_e$ decays with $s_H>(1.5\,{\rm GeV})^2$ is 15\%, 16\%,
and 15\%, respectively.  One should not conclude from the approximate agreement
between these models that the uncertainty in these predictions for the leakage
is less than a factor of two.

The larger the value of $\bar\Lambda$, the larger the fraction of $B\to
X_u\,e\,\bar\nu_e$ decays that leak out beyond $s_H=m_D^2$.  In Fig.~4 we plot
the model $\Phi\propto e^{-|\vec p\,|^2/p_F^2}$ with $m_{sp}=0.2\,$GeV and
$p_F$ taking three values ($p_F=0$, $0.3$, and $0.5\,$GeV) corresponding to the
choices $\bar\Lambda=0.2$, $0.4$ and $0.6\,$GeV.  For these values of $p_F$,
the model gives $\lambda_1=0$, $-0.14$, and $-0.38\,{\rm GeV}^2$ respectively,
in qualitative agreement with the correlation between $\bar\Lambda$ and
$\lambda_1$ from the electron energy spectrum in semileptonic $B$ decay
\cite{gremm,AM}.  The $p_F=0$ (dotted) curve is given analytically by the $n=0$
term in Eq.~(\ref{11}) with $\bar\Lambda=0.2\,$GeV.  The fraction of $B\to
X_u\,e\,\bar\nu_e$ decays with $s_H\geq m_D^2$ is 0, 2.8\%, and 12\%,
respectively.  The fraction of events with $s_H>(1.5\,{\rm GeV})^2$ is 0, 15\%,
and 31\%, respectively.  The rapid variation of these values with $\bar\Lambda$
shows again that a reliable determination of $|V_{ub}|$ from the number of
events in the region $s_H<\Delta^2$ is only possible if $\bar\Lambda$ does not
have too large a value.  This is especially true if experimental issues force
$\Delta^2$ to be significantly smaller then $m_D^2$.

\begin{figure}[t]
\centerline{\epsfysize=9truecm \epsfbox{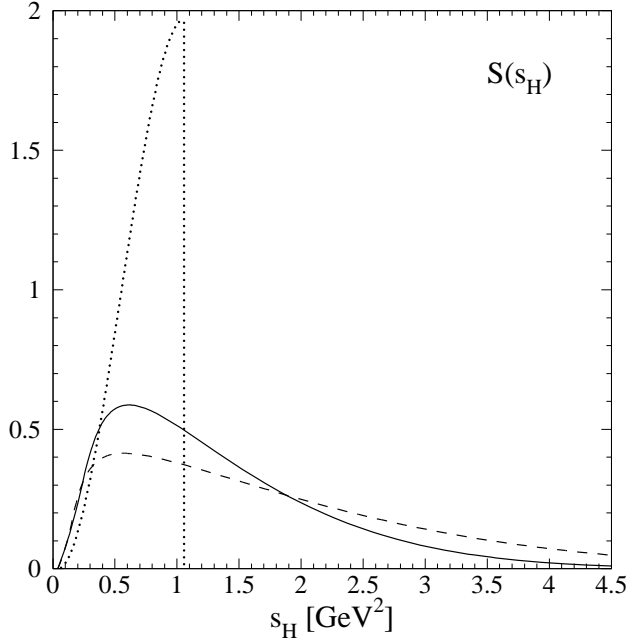}}
\caption[4]{The shape function $S(s_H)$ in the ACCMM model, with
$m_{sp}=0.2\,$GeV.  The dotted curve corresponds to $\bar\Lambda=0.2\,$GeV
($p_F=0$), the solid curve is $\bar\Lambda=0.4\,$GeV ($p_F=0.3\,$GeV), and
the dashed curve is $\bar\Lambda=0.6\,$GeV ($p_F=0.5\,$GeV).}
\end{figure}

We have investigated the utility of the hadronic invariant mass spectrum in
$B\to X_u\,e\,\bar\nu_e$ decay in the region $s_H<m_D^2$ for a possible
model-independent determination of $|V_{ub}|$.  Perturbative QCD corrections to
${\rm d}\Gamma/{\rm d}s_H$ of order $\alpha_s^2\beta_0$ were calculated paying
particular attention to kinematic effects arising from $\bar\Lambda$.  A
measurement of $\int_0^{\Delta^2}{\rm d}s_H\,({\rm d}\Gamma/{\rm d}s_H)$ can be
translated into a value of $|V_{ub}|$ using Fig.~2, and the result can then be
corrected (with some model dependence) for nonperturbative effects coming from
operators with dimension five and higher in the OPE.  If $\Delta^2\simeq m_D^2$
and $\bar\Lambda=0.4\pm0.1\,$GeV are experimentally feasible, then $|V_{ub}|$
can be extracted model-independently with about 10\% theoretical uncertainty
from the error in the value of $\bar\Lambda$.  In this case uncertainties
associated with higher dimension operators in the OPE are likely to be small. 
A determination of $\bar\Lambda$ with a precision of $\pm0.1\,$GeV from
experimental information on semileptonic $B$ decay and weak radiative $B$ decay
seems possible~\cite{gremm,AM,FLS,lbl1rad,lbl1else}.

It is possible that the experimental invariant mass resolution will necessitate
an upper cut on the hadronic invariant mass squared $\Delta^2$ which is
somewhat below $m_D^2$, in a region where nonperturbative effects that make the
spectrum leak beyond $\Delta^2$ are not negligible.  For
$s_H\sim\bar\Lambda\,m_B$, both nonperturbative strong interaction effects and
higher order perturbative corrections become important.  If $\Delta^2$ has to
be substantially smaller than $m_D^2$, then $\bar\Lambda$ cannot be too large
for this method of extracting $|V_{ub}|$ to remain viable.  In this case, the
theoretical uncertainty in $|V_{ub}|$ will depend sensitively on both the value
of $\Delta^2$ and the value and uncertainty in $\bar\Lambda$ at the time when a
measurement of $\int_0^{\Delta^2}{\rm d}s_H\,({\rm d}\Gamma/{\rm d}s_H)$ is
available.   In addition, since the nonperturbative effects introduce a certain
level of model-dependence, it will be important to compare the extracted value
of $|V_{ub}|$ from the hadronic invariant mass spectrum with its value from
other determinations, such as from exclusive decays~\cite{excl}.

\acknowledgements
While this work was being completed, a paper by Dikeman and Uraltsev  
appeared~\cite{DiUr} which addresses similar issues.  Z.L.~and M.B.W.~were  
supported in part by the Department of Energy under Grant  
No.~DE-FG03-92-ER40701.  A.F.F.~was supported in part by the National Science  
Foundation under Grant No.~PHY-9404057 and National Young Investigator Award  
No.~PHY-9457916, by the Department of Energy under Outstanding Junior  
Investigator Award No.~DE-FG02-94ER40869, and by the Alfred P.~Sloan  
Foundation.

{\tighten

}

\end{document}